\documentclass[preprint,floatfix,aps,prd,showpacs,footinbib,amsmath,amssymb,amsfonts,superscriptaddress]{revtex4-1}

\usepackage{graphicx}
\usepackage{color}
\usepackage{hyperref}
\usepackage{mathtools}
\usepackage{mathrsfs}
\usepackage{calrsfs}
\usepackage{comment}
\usepackage[utf8]{inputenc}
\usepackage{multirow} 
\usepackage{subfigure}
\expandafter\def\csname ver@subfig.sty\endcsname{}
\usepackage{varioref}

\expandafter\ifx\csname package@font\endcsname\relax\else
\expandafter\expandafter
\expandafter\usepackage
\expandafter\expandafter
\expandafter{\csname package@font\endcsname}%
\fi
\hypersetup{
	colorlinks=true,       
	linkcolor=blue,          
	citecolor=blue,        
}

\usepackage[section]{placeins}

\usepackage{fancybox}
\labelformat{figure}{Fig.~#1} 

\begin{document}
\title{Electromagnetic memory in arbitrary curved space-times}
	%
\author{Susmita Jana} 
\email{susmitajana@iitb.ac.in}
\affiliation{Department of Physics, Indian Institute of Technology Bombay, Mumbai 400076, India}
\author{S. Shankaranarayanan}
\email{shanki@iitb.ac.in}
\affiliation{Department of Physics, Indian Institute of Technology Bombay, Mumbai 400076, India}
\begin{abstract}
The gravitational memory effect and its electromagnetic (EM) analog are potential probes in the strong gravity regime. In the literature, this effect is derived for static observers at asymptotic infinity. While this is a physically consistent approach, it restricts the space-time geometries for which one can obtain the EM memory effect. To circumvent this, we evaluate the EM memory effect for comoving observers (defined by the 4-velocity $u_{\mu}$) in arbitrary curved space-times. Using the covariant approach, we split Maxwell's equations into two parts --- projected parallel to the 4-velocity $u_{\mu}$ and into the 3-space orthogonal to $u_{\mu}$. Further splitting the equations into $1+1+2$-form, 
we obtain the acceleration vector of the comoving observer located in a 2-D surface orthogonal to the direction of propagation of the EM waves. We refer to this expression as the \emph{master equation} for the EM memory in an arbitrary curved space-time. The master equation corresponding to the acceleration of the comoving observer in the 2-D surface provides a physical understanding of the contribution to the EM memory. For instance, the leading order contribution only requires information about the total energy density of the EM field, while the subleading contributions contain information about the space-time geometry and the other components of the energy-momentum tensor of the EM field. To our knowledge, this is the first time a transparent and easily applicable final expression for electromagnetic memory has been derived for a general curved space-time. We then obtain EM memory for specific space-time geometries and demonstrate the advantages of our approach.

%
\end{abstract}
\maketitle
%

%
\section{Introduction}
\label{sec:introduction}

LIGO-VIRGO-KAGRA has detected close to 100 gravitational wave (GW) sources. GW signals emanating from a black hole or neutron star binaries have opened many new research avenues in astronomy, cosmology, and fundamental physics~\cite{2009-Sathyaprakash.Schutz-LRR,Arun:2013bp,2019-Barack.etal-CQG,2022-Shanki.Joseph-GRG}. GWs provide a unique way to test gravity's most extreme, non-linear regime in novel ways. The planned third-generation ground-based detector (Cosmic Explorer and the Einstein Telescope) will allow us to peer far deeper, and LISA will open a new observational window at low frequencies. With more sensitive detectors shortly, the focus has been to understand the physical effects of GWs. 
\emph{Gravitational wave memory} is one such effect~\cite{Zeldovich:1974gvh, PhysRevD.15.2069, kovacs1978generation,  Braginsky:1986ia, 1987-ThorneBraginskii-Nature,PhysRevD.45.520, PhysRevD.44.R2945,2010-Favata-CQG,2014-Bieri.Garfinkle-PRD}.

GW memory effects --- physically observable phenomena that modify the state of gravitational-wave detectors a little bit from their original undisturbed state --- are one of the key predictions of general relativity
~\cite{PhysRevD.15.2069, kovacs1978generation, 1987-ThorneBraginskii-Nature,1991-Christodoulou-PRL}. 
GW memory effects can be divided into two types~\cite{2010-Favata-CQG,2014-Bieri.Garfinkle-PRD}: \emph{null memory} that occurs when radiation or massless particles escape from a system to null infinity, and \emph{ordinary memory} that occurs when the detector recoils relative to its initial center of mass frame. The GW memory is characterized as a gravitational wave signal approaching a nonzero finite value. This aspect of the GW signal is yet to be observed, although LISA is predicted to observe it~\cite{2009-Favata-ApJL}.

\begin{figure}[!htb]
\includegraphics[width=0.75\textwidth]{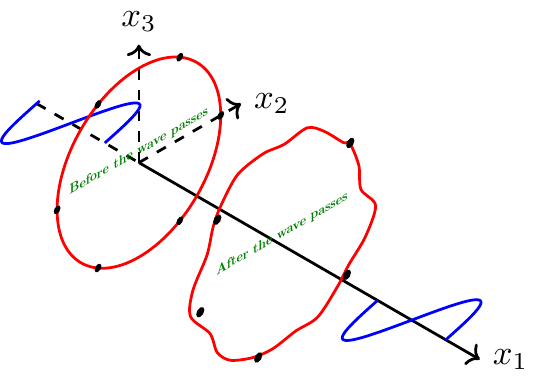}
\caption{Electromagnetic memory effect that lies in the 2-D surface orthogonal to the direction of the coming wave.}
\label{fig:tikz:EM-Memory}
\end{figure}

Recently, it has been realized that the memory effect can be thought of as a vacuum transition between two different states related by an asymptotic transformation~
\cite{2016-Strominger.Zhiboedov-JHEP,2017-Zhang.etal-PRD}. Since such asymptotic transformations also occur for other gauge theories, there has been an intense activity to obtain analogous memory effects in other gauge theories~\cite{Bieri:2013hqa,Winicour:2014ska,2014-Strominger-JHEP,2017-Pate.etal-PRL,2019-Satishchandran.Wald-PRD}. 
Since electromagnetic (EM) theory is the simplest of all gauge theories and can be a potential probe, \emph{electromagnetic memory} has received much attention~\cite{2012-Bieri-CQG,2015-Susskind-hep-th,2017-Gomez.Raoul-JHEP,2017-Hamada-JHEP,2018-Hamada.Sotaro-JHEP,2019-Mao-PRD,2019-Heissenberg-hep-th,2019-Jokela.Sarkkinen-PRD,2020-Aldi.Maurizio-PLB,2020-Hirai.Sotaro-JHEP,2020-Mao.Tan-PRD,2022-Taghiloo-hep-th,2022-AtulBhatkar-PRD,2022-Seraj-hep-th}. 
Like in GW memory, an EM wave generates a permanent change in the relative velocity of test-charged particles attached to a  detector in the 2-D surface perpendicular to the direction of propagation of the wave while passing through the detector~[cf. (\ref{fig:tikz:EM-Memory})].
In other words, EM waves directly displace test particles by giving them a momentum (kick), resulting in a relative velocity change.  
This is different from GW memory as the GW does not displace test particles. Instead, GW distorts the space-time geometry itself, which causes a change in separation between two test particles.  

Bieri and Garfinkle were the first to propose the memory effect due to electromagnetic waves~\cite{Bieri:2013hqa}. Like in GW memory, they showed that EM waves produce two types of momentum kicks.  
In Ref. \cite{Winicour:2014ska}, Winicour showed the absence of memory effect generated by the electromagnetic field coming from distant sources for a bound charge distribution and the non-existence of memory effect due to the magnetic field. 

In the case of GW memory, gravitational radiation must reach the detector. Likewise, EM radiation also has to reach null infinity to generate \emph{null kick} memory. Hence to calculate EM memory, one needs to know the properties of the electric field and radiation at null infinity~\cite{Bieri:2013hqa}.
More specifically, the original approach by Bieri and Garfinkle requires prior knowledge about the behavior of the fields in asymptotic limits. It can be extended to conformally flat space-times ~\cite{2022-AtulBhatkar-PRD, 2022-EnriquezRojo-JHEP}. Also, the analysis does not provide any physical understanding of why the EM memory has such a form in flat and conformally flat space-times.

This leads us to the following questions: Can we 
derive a master equation for \emph{EM memory} in a generic curved space-time? What role does curved geometry play in EM memory? Can we have a physical understanding of the various contributions to EM memory? This work addresses these three questions using $1+3$ covariant formalism~\cite{1955-Heckmann-zap,1955-A.K.Raychaudhuri-PRL,1996-vanElst.Ellis-CQG,1998-Ellis-NATOSci,2008-Tsagas.etal-PRep,2012-ellis_maartens_maccallum-Book}.

There are two reasons why covariant formalism is better suited to studying EM memory. First, as mentioned earlier, when the EM wave propagates in a given spatial direction, the net momentum experienced by the particle lies in the $2$-D surface orthogonal to the direction of propagation of the EM wave (for a pictorial representation, see \ref{fig:tikz:EM-Memory}). In other words, the EM memory affects the test particle lying on the 2-D surface. Hence, it is more natural to have a formalism that identifies such a dynamical 2-D surface and evaluates EM memory. Second, like in fluid mechanics, we can observe the flow of EM radiation in two ways. First, as in Refs.~\cite{Bieri:2013hqa,Winicour:2014ska}, an asymptotic stationary observer monitors changes in Electric and Magnetic fields of the incoming EM radiation. Second, a comoving observer monitors changes in Electric and Magnetic fields. In fluid mechanics, these are referred to as the Lagrangian and Lagrangian descriptions of flow, respectively. It is well-known that the Lagrangian description is better suited for fluids and in cosmology~\cite{1996-vanElst.Ellis-CQG,1998-Ellis-NATOSci,2012-ellis_maartens_maccallum-Book}.

In this work, we evaluate the memory effect 
using the $1+1+2$ {covariant} formalism~\cite{1996-vanElst.Ellis-CQG,2003-Clarkson.Barrett-CQG,2004-Clarkson.P.K.S-APJ,2005-Tsagas-CQG,2021-Mavrogiannis.Tsagas-PRD}. The $1 + 1 + 2$ decomposition of space-time is a natural extension of the $1 + 3$ formalism in which the three-space is further decomposed to a given spatial direction. This approach is also referred to as \emph{semi-tetrad formalism}~\cite{2017-Sayuri.Sunil.Maharaj-PRD, 2019-Sayuri.Sunil.Maharaj-JMP, 2020-Hansraj.Sunil.Maharaj-GRG, 2022-Singh-Sunil.Maharaj, 2020-Khambule.SunilMaharaj-CQG}. The principle advantage is that we can evaluate the net momentum (kick) vector on the 2-D surface for arbitrary space-time. Since this affects all the test particles on the 2-D surface, we refer to this as \emph{memory vector}. This can also be understood using the fact that the electric and magnetic fields are transverse to the direction of propagation of the EM wave. Further splitting the equations into $1+1+2$-form, 
we obtain the acceleration vector of the comoving observer located in a 2-D   surface orthogonal to the direction of propagation of the EM waves. We refer to this expression as the \emph{master equation} for the EM memory in an arbitrary curved space-time. The master equation corresponding to the acceleration of the comoving observer in the 2-D surface  provides a physical understanding of the contribution to the EM memory. For instance, the leading order contribution only requires information about the total energy density of the EM field, while the subleading contributions contain information about the space-time geometry and the other components of the energy-momentum tensor of the EM field. To our knowledge, this is the first time a transparent and easily applicable final expression for electromagnetic memory has been derived for a general curved space-time. We then obtain EM memory for specific space-time geometries and demonstrate the advantages of our approach.


The rest of this work is organized as follows: In Sec.~\ref{sec:covariant_formalism}, we provide an overview of the two --- $1 + 3$ and $1 + 1 + 2$ --- covariant formalisms and obtain the key geometrical quantities. Then, in Sec.~\ref{sec:EMCovariant}, we rewrite Maxwell's equation in $1 + 3$ and $1 + 1 + 2$ covariant formalisms in arbitrary space-time. Next, in Sec.~\ref{sec:memory_generic}, we obtain the master equation for the EM memory 
in arbitrary space-time and discuss the key features. In Sec.~\ref{sec:memory_specific_sptm}, we then obtain EM memory for specific space-times and compare them with the known results in the literature. Finally, in Sec.~\ref{sec:conclusion}, we summarise our results and discuss possible future directions.

In this work, we use $(-, +, +, +)$ metric signature and set $c = 1/(4 \pi \epsilon_0) = 1$.  A dot denotes a derivative with respect to the proper time $\tau$. A prime denote derivative w.r.t the space-like vector $n^{\mu}$. For easy comparison, we follow the notations of   Ref.~\cite{2012-ellis_maartens_maccallum-Book}.


%
%
%
%

\section{Overview of Covariant formalism}
\label{sec:covariant_formalism}
A covariant theory like general relativity does not favor any particular coordinates. However, splitting tensors into time and spatial parts is typically required for its physical meaning. Thus, the splitting is achieved by rewriting Einstein's equations as a set of constraint and evolution equations in a three-dimensional framework. This allows for an intuitive evaluation of the relevant physical system. 

A choice of coordinates defines a
threading of space-time into lines and slicing into hypersurfaces~\cite{1994-Boersma-Dray-GRG}. Thus, the splitting procedure can be carried out in two distinct ways: First, by employing the so-called $(3+1)-$ formalism or slicing of space-time~\cite{2008-Alcubierre-Book}. Second, by employing $(1+3)-$ formalism, or threading of space-time~\cite{1996-vanElst.Ellis-CQG,1998-Ellis-NATOSci,2012-ellis_maartens_maccallum-Book}. In the $(3+1)-$ decomposition, the time is a label of space-like slices $\Sigma_t$ with space coordinates $x_i$. In contrast, in the $(1+3)-$ splitting, the time-like world lines have coordinate $\tau$ and are labeled by $x^{\mu}$. In the $(3+1)-$ formulation, the construction only requires space-like hypersurfaces and does not demand causality of the time curves. However, in the $(1+3)-$ approach, every tensor is split into the parallel and orthogonal directions to a time-like vector (curves). Furthermore, it does not provide any condition on the causality of the spatial distances. Though the two approaches provide different points of view, it has been shown that they are equivalent for space-times with symmetries~\cite{1994-Boersma-Dray-GRG}. We use the covariant $1+3$ formalism in this work to obtain EM memory. As mentioned in the introduction, covariant formalism provides a physical understanding of the origin of EM memory in arbitrary space-time.

\subsection{Covariant 1+3 Formalism}
\label{subsec:1+3}
\begin{figure}[!htb]
\includegraphics[width=0.5\textwidth]{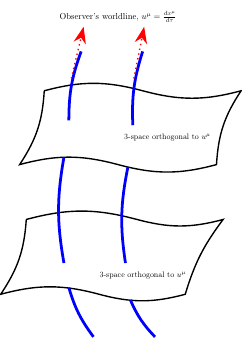}
\caption{Visualisation of $1+3$ formalism.}
\label{fig:3+1}
\end{figure}
Heckmann, Schucking, and Raychaudhuri developed the covariant approach to General relativity in the 1950s~\cite{1955-Heckmann-zap, 1955-A.K.Raychaudhuri-PRL} and was later used in different gravitational and cosmological models~\cite{1996-vanElst.Ellis-CQG,1998-Ellis-NATOSci,2008-Tsagas.etal-PRep,2012-ellis_maartens_maccallum-Book}. To decompose the 4-D space-time in $(1+3)-$ formalism, we introduce a  family of observers with worldlines tangent to a timelike $4$-velocity vector $u^{\mu}$ satisfy the following:
\begin{align}
\label{eq:4-velocity}
u^{\mu} = \frac{dx^{\mu}}{d\tau};
\quad
u^{\mu}u_{\mu} = -1 \, ,
\end{align}
where $\tau$ is the proper time measured along the fundamental world line. See \ref{fig:3+1}. Using the $4$-velocity ($u^{\mu}$) we can define the following projection tensors~\cite{2012-ellis_maartens_maccallum-Book,1998-Ellis-NATOSci}:
\begin{subequations}
\begin{align}
\label{eq:projection1-1+3}
& U^{\mu}\,_{\nu} = - u^{\mu}u_{\nu}; \quad U^{\mu}\,_{\nu}\, U^{\nu}\,_{\gamma} = U^{\mu}\,_{\gamma}; \quad U^{\mu}\,_{\mu} = 1\\
\label{eq:projection2-1+3}
& h_{\mu\nu} = g_{\mu\nu} + u_{\mu}u_{\nu}; \quad h^{\mu}\,_{\nu}\, h^{\nu}\,_{\gamma} = h^{\mu}\,_{\gamma}; \quad h^{\mu}\,_{\mu} = 3; \quad h_{\mu\nu} \, u^{\nu} = 0
\end{align}
\end{subequations}
$u^{\mu}$, and hence $U^{\mu}\,_{\nu}$, projects physical quantities parallel to the $4$-velocity of the observer and $h_{\mu\nu}$ projects quantities into the $3$-space orthogonal to $u^{\mu}$. The tensor
$h_{\mu\nu}$ provides the metric properties of the instantaneous $3$-space as well in the absence of rotation or vorticity.

In this formalism, the projection of the vector ($V^{\nu}$) orthogonal to $u^{\mu}$ is defined as $ V_{<\mu>}$. Similarly, the trace-less part of a rank-$2$ tensor ($S^{\alpha\beta}$) projected into space orthogonal to $u^{\mu}$ is defined as $S_{<\mu\nu>}$. Mathematically, these are given by:
\begin{align}
    V_{<\mu>} := h_{\mu\nu}\, V^{\nu}; \hspace{0.5cm} S_{<\mu\nu>} := \left(h_{\mu\alpha}h_{\nu\beta} - \frac{1}{3} h_{\mu\nu}h_{\alpha\beta}\right) \, S^{\alpha\beta}
\end{align}
The projection of the time derivative and orthogonal spatial derivative of any vector ($V^{\nu}$) and tensor ($S^{\alpha\beta}$) are defined as:
\begin{align}
     \dot{ V}^{<\mu>} := h^{\mu}\,_{\alpha}u^{\nu}\nabla_{\nu}\, V^{\alpha} ; \hspace{0.5cm} D_{\alpha}\, S^{\beta\gamma}  := h^{\mu}\,_{\alpha}\, h^{\beta}\,_{\nu} \,h^{\gamma}\,_{\rho} \,\nabla_{\mu}\, S^{\nu\rho}
\end{align}
%
%
%
The covariant derivative of $u_{\mu}$ can be split into two parts: $1)$ directional derivative along the tangent to the world line, $2)$ spatial derivative in the 3-space orthogonal to $u^{\nu}$. This can further be split into trace, traceless symmetric and anti-symmetric tensor: 
%
%
%
\begin{align}
\label{eq:cov_derivative_u_final}
\nabla_{\nu}u_{\mu} 
= \frac{\Theta}{3} h_{\mu\nu}  + \sigma_{\mu\nu} +  \omega_{\mu\nu} - \dot{u}_{\mu}u_{\nu} \, .
\end{align}
In the above equation, $\sigma_{\mu\nu}$ is the symmetric expansion tensor that describes the distortion in the matter flow, $\Theta$ corresponds to the expansion rate of the matter w.r.t the observer, $\omega_{\mu\nu}$ is the anti-symmetric vorticity tensor describing the rotation of the matter w.r.t a non-rotating frame. The last term refers to the relativistic acceleration vector (the directional derivative) $\dot{u}_{\mu} = u^{\nu}\nabla_{\nu}$ which corresponds to the degree to which the matter moves under forces other than gravity plus inertia. Further, using the vorticity tensor, we can define the following quantity called the vorticity vector:
\begin{align}
\label{eq:vorticity_vec}
 & \omega^{\nu} = -\frac{1}{2}  \epsilon^{\mu\nu\alpha\beta}\omega_{\alpha\beta}\, u_{\mu} 
\end{align}
where, $\epsilon^{\mu\nu\alpha\beta} = \frac{1}{\sqrt{-g}} \eta^{\mu \nu \rho \sigma}$ is fully antisymmetric tensor, $\eta^{\mu \nu \rho \sigma}$ is Levi-Civita symbol whose values are $\pm 1$ and we set $\eta^{0123}=1=-\eta_{0123}$~\cite{1971-Ellis-Proc}. 
The Levi-Civita $3$-tensor is defined as:
\begin{align}
\label{eq:3-LeviCivita}
   \epsilon_{\mu\nu\alpha}\equiv \epsilon_{\mu\nu\alpha\beta}u^{\beta} \, ,
\end{align}
and satisfies the following relations: $\epsilon_{\mu\nu}u^{\nu} = 0 $ and $\epsilon^{\mu\nu\alpha\beta} = 2\left( \, u^{[\mu} 
\epsilon^{\nu]\alpha\beta} - 
\epsilon^{\mu\nu[\alpha}u^{\beta]} \, \right)$. The square bracket w.r.t the indices refers to antisymmetrization.

\subsection{1+1+2 covariant formalism}
\label{subsec:1+1+2}

The $1+3$-\textit{covariant} formalism is well-suited for relativistic cosmology because, at the largest observable scales, the universe is homogeneous and isotropic~\cite{1998-Ellis-NATOSci}. These symmetries allow the slicing or threading of the 4-D space-time manifold into a one-parameter family of spacelike hypersurfaces corresponding to cosmic time. 
Interestingly, it is easy to show that 
in the Friedmann-Lemaitre-Robertson-Walker (FLRW) background, all physical quantities except for the volume expansion $\Theta$ and the energy density vanish. Using the Stewart-Walker lemma, in this formalism, it was possible to construct gauge invariant quantities up to second order in cosmological perturbations~\cite{1989-Ellis.Bruni-PRD, 1989-Ellis.Hwang.Bruni-PRD}. However, the $1+3$-formalism is not suited if the space-time is inhomogeneous, like spherical symmetry or space-times with local rotational symmetry (LRS)~\cite{2003-Clarkson.Barrett-CQG}. In such cases, 
splitting the $3$-space orthogonal to the time-like congruence into one spacelike direction and a $2$-space is apt~\cite{1996-vanElst.Ellis-CQG}. Thus, the $1 + 1 + 2$ decomposition of space-time is a natural extension of the $1 + 3$ formalism
in which the three-space is further decomposed to a given spatial direction. This approach is called semi-tetrad formalism~\cite{2017-Sayuri.Sunil.Maharaj-PRD, 2019-Sayuri.Sunil.Maharaj-JMP, 2020-Hansraj.Sunil.Maharaj-GRG, 2022-Singh-Sunil.Maharaj, 2020-Khambule.SunilMaharaj-CQG}.

As mentioned in the Introduction, our interest is to evaluate the net momentum experienced by a test particle after the electromagnetic wave passes through the space-time point. In the covariant $1+3$ formalism, the test 
particle is the fundamental time-like observer. As depicted in \eqref{fig:tikz:EM-Memory}, when the EM wave propagates in a given spatial direction, the net momentum experienced by the particle lies in the $2$-D surface orthogonal to the direction of propagation of the EM wave. In other words, the net momentum (kick) vector lies in the $2$-D surface. Thus, the net memory effect of the test particle will lie on the 2-D surface; hence, we will refer to this as the \emph{memory vector}. This can also be understood using the fact that the electric and magnetic fields are transverse to the direction of propagation of the EM wave. Thus, it is cogent to further split the 3-space to $1+2$-space. 

More specifically, choosing a generic space-like vector ($n^{\mu}$), we split the $3$-space into 1 + 2-space~\cite{2003-Clarkson.Barrett-CQG,2004-Clarkson.P.K.S-APJ,2005-Tsagas-CQG,2021-Mavrogiannis.Tsagas-PRD}. The space-like vector ($n^{\mu}$) satisfies the following conditions:
\[
n^{\mu}n_{\mu} = 1, \quad n^{\mu}u_{\mu} = 0 \, . 
\]
Like in the $1 + 3$-formalism, we project the vectors and tensors defined in $3$-space along the space-like direction ($n^{\mu}$) and into the $2$-space that is orthogonal to $n^{\mu}$. Here again, the projection tensor ($\Tilde{h}_{\mu\nu}$) need to be defined: 
\begin{align}
\label{eq:projection-1+1+2}
\Tilde{h}_{\mu\nu} = h_{\mu\nu} - n_{\mu}n_{\nu}; \quad \Tilde{h}^{\mu}\,_{\nu}\, \Tilde{h}^{\nu}\,_{\gamma} = \Tilde{h}^{\mu}\,_{\gamma}; \quad \Tilde{h}^{\mu}\,_{\mu} = 2; \quad \Tilde{h}_{\mu\nu} \, u^{\nu} = 0; \quad \Tilde{h}_{\mu\nu} \, n^{\nu} = 0 \, .
\end{align}
All the vectors and tensors defined in the $3$-space in the $1 + 3$-formalism can be split into $1+2$ form. For instance, an arbitrary space-like vector $V^{\mu}$ (defined in the $3$-space) can be written as:
\begin{align}
\label{eq:vector-1+2}
V^{\mu} = \mathcal{V} n^{\mu}+\mathcal{V}^{\mu}
\end{align}
where, $\mathcal{V} = V^{\mu}n_{\mu} $ and $\mathcal{V}^{\mu} = \Tilde{h}^{\mu}\,_{\nu}V^{\nu}$. Similarly an arbitrary tensor $v_{\mu\nu}$ on the 3-space can be split as:
\begin{align}
\label{eq:tensor-1+2} 
v_{\mu\nu}=V\left(n_{\mu} n_{\nu}-\frac{1}{2} \tilde{h}_{\mu\nu}\right)+2 V_{(\mu} n_{\nu)}+V_{\mu\nu} \, ,
\end{align}
where $V_{(\mu} n_{\nu)} = (V_{\mu} n_{\nu} +  n_{\nu} V_{\mu})/2$.  
Similarly, the relative acceleration of the time-like observer and other geometrical quantities  defined in $3$-space can be written in $1+2$ space as:
\begin{align}
\label{eq:udot_1+2}
&\dot{u}^{\mu}=\mathcal{A} n^{\mu}+\mathcal{A}^{\mu} \\
\label{eq:ndot_1+2}
&\dot{n}^{\mu}=\mathcal{A} u^{\mu}+\alpha^{\mu} \\
\label{eq:omega_1+2}
&\omega^{\mu}=\Omega n^{\mu}+\Omega^{\mu} \\
\label{eq:sigma_1+2}
&\sigma_{\mu\nu}=\Sigma\left(n_{\mu} n_{\nu}-\frac{1}{2} \tilde{h}_{\mu\nu}\right)+2 \Sigma_{(\mu} n_{\nu)}+\Sigma_{\mu\nu} 
\end{align}
where $\Dot{n}^{\mu} := u_{\nu}\nabla_{\nu}\ n^{\mu}$ is the relative acceleration of the space-like vector along the time-like observer. Here, $\mathcal{A}^{\mu}, \alpha^{\mu}, \Sigma_{\mu\nu},\, \Omega^{\mu}$ are orthogonal to $n^{\mu}$ as well as $u^{\mu}$. Also, $\mathcal{A}^{\mu}, \Omega^{\mu} (\Sigma_{\mu\nu})$ are the vectors (tensor) projected on the 2-space.  In this formalism, we define the alternating Levi-Civita $2$-tensor 
\begin{align}
   \label{eq:2-LeviCivita}
   \epsilon_{\mu\nu}\equiv \epsilon_{\mu\nu\alpha}n^{\alpha}
\end{align}
which is orthogonal to $n^{\mu}$ and has components only in the $2$-space. {Given an arbitrary vector $V^{\mu}$ in the $2$-space, we can construct another vector $\epsilon_{\mu\nu}V^{\nu}$ that is orthogonal to $V^{\mu}$ which is in the 2-space and has the same length.}

The $1 + 2$ splitting of the 3-space leads to a new directional derivative along the space-like vector $n^{\mu}$:
\begin{align}
\label{eq:derivative-n}
& v_{\mu\nu}^{\prime} \equiv n^{\alpha}D_{\alpha} v_{\mu\nu}\\
\label{eq:covDerivative-2space}
& \Tilde{D}_{\alpha} v_{\mu\nu} \equiv \Tilde{h}_{\alpha}\,^{\beta}\Tilde{h}_{\mu}\,^{\rho}\Tilde{h}_{\nu}\,^{\sigma} D_{\beta} v_{\rho\sigma} \, .
 \end{align}
The derivative in Eq. \eqref{eq:derivative-n} physically correspond to the variation of the physical quantities on the 2-space along the space-like vector $n^{\mu}$. The derivative ($\Tilde{D}$) in Eq.~\eqref{eq:covDerivative-2space} corresponds to the variation of the physical quantities that lie in the $2$-space.
These will contribute to the memory vector.

As we split the covariant derivative of $u_{\mu}$ in Eq.~\eqref{eq:cov_derivative_u_final}, similarly we can split the covariant derivative of $n_{\mu}$ as:
\begin{align}
\label{eq:cov_derivative_n} 
D_{\nu}n_{\mu} = \Tilde{D}_{\nu}n_{\mu}+n_{\mu}n_{\nu}^{\prime}=\Tilde{\sigma}_{\mu\nu} + \Tilde{\omega}_{\mu\nu}+ \frac{1}{2}\Tilde{\Theta}\Tilde{h}_{\mu\nu} + n_{\mu}n_{\nu}^{\prime}
\end{align}
where, $ \Tilde{\sigma}_{\mu\nu}\equiv \Tilde{D}_{<\nu}n_{\mu>}$, $ \Tilde{\omega}_{\mu\nu}\equiv \Tilde{D}_{(\nu}n_{\mu)}$ and $ \Tilde{\Theta} = \Tilde{D}^{\mu}n_{\mu}$ are shear, vorticity and the surface expansion-contraction scalar respectively and $n_{\mu}^{'}$ is the spatial derivative along $n^{\mu}$. Thus, $\Tilde{D}_{\nu}n_{\mu}$ describes the kinematic properties or the relative motion of the space-like curves in the $2$-surface orthogonal to $n^{\mu}$. 
We can obtain the relation between the kinematic quantities derived from the motion of time-like vector $u_{\mu}$ and kinematic quantities in $2$-space derived from the space-like vector $n^{\mu}$. See, for instance, Ref.~\cite{2021-Mavrogiannis.Tsagas-PRD}.

\section{Electromagnetic theory in covariant formalism}
\label{sec:EMCovariant}

The covariant formalism has been extensively employed in studying the evolution of electromagnetic fields in curved space-time~\cite{2005-Tsagas-CQG}. 
In the covariant formulation, the dynamics and kinematics are constricted by the Bianchi and Ricci identities. The $(1 + 3)-$ covariant formulation permits the classification of cosmological models, a fluid description of the matter field in FLRW universes. However, as mentioned earlier, the $1+3$-formalism is not suited if the space-time is inhomogeneous, like spherical symmetry or space-times with LRS~\cite{2003-Clarkson.Barrett-CQG}. In such cases, 
the $1+1+2$-\textit{covariant} or semi-triad formalism are better suited. 

Since we aim to derive EM memory for arbitrary space-times, we use $1+1+2$-covariant formalism. We obtain a  
generic form of the EM memory effect by evaluating the change in the velocity vector $\Delta u^{\mu}$  that 
lie in the $2$-space. In order to do so, we fix the space-like direction to be the direction of the propagation of the wave. In the case of spherically symmetric space-time, this naturally translates to the radial direction. One key advantage is that the electromagnetic theory in the $1+1+2$ formalism helps to understand the evolution and dynamics of the EM fields along the space-like direction and in the $2$-space normal to $n^{\mu}$ and $u^{\mu}$. Our approach makes geometrical contributions to the memory effect more transparent.

In the next subsection, we rewrite Maxwell's equations in $1 + 3$ formalism in an arbitrary space-time. Later, we formulate the evolution equations of the EM fields in the $2$-space and two constraint equations of the same along $u^{\mu}$ and $n^{\mu}$~\cite{2021-Mavrogiannis.Tsagas-PRD}. The key advantage is that we can obtain the memory vector from the projected acceleration vector onto the 
$2$-space.


\subsection{In 1+3 formalism}
\label{subsec:Maxwell_1+3}

The fundamental objects are the Maxwell electromagnetic field tensor $F^{\mu\nu}$. The $(1+3)$ covariant formalism of Maxwell's electromagnetic theory provides a way to study the interaction of EM fields with different components of general space-time geometry~\cite{2005-Tsagas-CQG}.
With the $(1+3)$ decomposition, it is possible to split $F^{\mu\nu}$ into the electric and magnetic fields. Note that the local coordinates are mathematical parameters that label the points of the space-time manifold $M$; therefore, the electric and magnetic fields may not have a direct physical meaning. In order to make measurements, an observer brings in an additional structure on $M$ by introducing the orthonormal coframe field. This gives rise to the split of Maxwell's tensor $F$ into the physical electric and magnetic fields. 

Specifically, formalism allows us to split the equations of motion of the fields and currents into two parts: 
\begin{enumerate}
    \item projected parallel to the $4$-velocity $u^{\mu}$ of the fundamental observer 
    \item projected into the $3$-space orthogonal to $u^{\mu}$.
\end{enumerate}
To keep the calculations tractable, we 
perform all the calculations in source-free and lossless regions. However, the EM memory analysis can be straightforwardly extended to these regions.  In the source-free regions, Maxwell's equations are:
\begin{align}
\label{eq:Maxwell_Eq1}
& \nabla_{\nu}F^{\mu\nu} = 0\\
\label{eq:Maxwell_Eq2}
& \nabla_{[\gamma}F_{\mu\nu]} = 0; \quad \text{or} \quad  \nabla_{\nu}{F^*}^{\mu\nu} = 0 \, ,
\end{align}
where ${F^*}^{\mu\nu}$ is the dual to $F^{\mu\nu}$ and is defined as ${F^*}^{\mu\nu} = (1/2) \epsilon^{\mu\nu\alpha\beta} F_{\alpha\beta}$.

In the $1+3$ formalism, by projecting $F^{\mu\nu}$ and ${F^*}^{\mu\nu}$ along the time-like $4$-velocity vector, 
we can decompose them into electric and magnetic parts. The electric ($E^{\mu}$) and magnetic ($B^{\mu}$) 4-vectors are defined as: 
\begin{align}
\label{eq:Maxwell_E_field}
 E^{\mu}:= F^{\mu\nu} u_{\nu} \\ 
\label{eq:Maxwell_B_field}
B^{\mu} :=  {F^*}^{\mu\nu} u_{\nu} 
\end{align}
From the above definitions, we infer:
\begin{align}
\label{eq:orthoprop_Maxwell_E_B}
E^{\mu}u_{\mu}=0; \quad B^{\mu}u_{\mu}=0
\end{align}
which implies $E^{\mu}$ and $B^{\mu}$ have only spatial components. Given this, we can rewrite $F_{\mu\nu}$ and ${F^*}^{\mu\nu}$ as:
\begin{align}
\label{eq:Maxwell-define_F}
F_{\mu\nu} & = u_{\mu}E_{\nu} - u_{\nu}E_{\mu} + \epsilon_{\mu\nu\alpha\beta} B^{\alpha}u^{\beta} \\
\label{eq:Maxwell-define_dual_F}
\tilde{F}^{\alpha\beta} 
& = \epsilon^{\alpha\beta\mu\nu} u_{\mu}E_{\nu} + \left(\, u^{\alpha}B^{\beta}  - u^{\beta}B^{\alpha} \, \right) \, .
\end{align}
From the above expressions, we see that 
the simultaneous transformations $E^{\mu} \rightarrow - B^{\mu}$, $B^{\mu} \rightarrow  E^{\mu}$ leads to ${F^*}^{\mu\nu} \rightarrow F^{\mu\nu}$. This implies that 
we can obtain the second Maxwell's equation \eqref{eq:Maxwell_Eq2} from the first Maxwell's equation \eqref{eq:Maxwell_Eq1} or vice-versa. 
More specifically, if we obtain the time-like part and space-like part of Maxwell's equations \eqref{eq:Maxwell_Eq2}, we can write the time-like part and space-like part of the other Maxwell's equations \eqref{eq:Maxwell_Eq1} by substituting $E^{\mu} \rightarrow - B^{\mu}$, $B^{\mu} \rightarrow  E^{\mu}$.

%

In the rest of this subsection, we obtain Maxwell's equations by projecting along 
$u_{\mu}$ (time-like part) and $h_{\mu\nu}$ (space-like part)~\cite{2009-Subramanian-AsNa}.
We first obtain the time-like part of  Eq.~\eqref{eq:Maxwell_Eq2} by multiplying it with $u_{\mu}$: 
%
\begin{align}
\label{eq:Maxwell_Eq2_timelike}
u_{\alpha} \left(\, \nabla_{\beta}\tilde{F}^{\alpha\beta} \, \right)= 0 
\end{align}
Using the decomposition in Eq.~\eqref{eq:Maxwell-define_dual_F}, the above expression becomes:
\begin{align}
\label{eq:Maxwell_Eq2_timelike_Simp}
\nabla_{\beta}B^{\beta} - B^{\beta} \dot{u}_{\beta} + \left(\nabla_{\beta} u_{\alpha}\right) \, \epsilon^{\alpha\beta\mu\nu} u_{\mu}E_{\nu} = 0 
\end{align}
We simplify the above equation using the following steps: First, we combine the first two terms in the LHS. From Eq.~\eqref{eq:Maxwell_Eq2_timelike}, we have
$B^{\beta} \dot{u}_{\beta} = - u_{\beta}  \dot{B}^{\beta} = - u_{\beta} u^{\alpha}\nabla_{\alpha} {B}^{\beta}$. Substituting in the second term of the above expression, we have $ \delta_{\beta}^{\alpha}\, \nabla_{\alpha}B^{\beta} + u_{\beta} u^{\alpha}\nabla_{\alpha} {B}^{\beta} = h^{\alpha}_{\beta} \left(\nabla_{\alpha}B^{\beta}\right) $. 
Substituting $\nabla_{\beta} u_{\alpha}$ from 
Eq.~\eqref{eq:cov_derivative_u_final} and using the definition of vorticity vector in Eq.~\eqref{eq:vorticity_vec}, the third term in the LHS of the above expression simplifies to 
$- 2\omega^{\beta}E_{\beta}$. Thus, the time-like part of Eq.~\eqref{eq:Maxwell_Eq2} reduces to:
\begin{align}
\label{eq:Maxwell_Eq2_timelike_Final}
D_{\beta}B^{\beta} = 2 \omega^{\beta}E_{\beta}  \, .
\end{align}

The space-like part of  Eq.~\eqref{eq:Maxwell_Eq2} can be obtained by multiplying it with $h_{\mu}\,^{\nu}$,
\begin{align}
\label{eq:Maxwell_Eq2_spacelike}
h_{\alpha}\,^{\rho} \left(\, \nabla_{\beta}\tilde{F}^{\alpha\beta} \, \right)= 0 
\end{align}
Using a series of steps, the above expression can be rewritten as:
\begin{align}
\label{eq:Maxwell_Eq2_spacelike_Final}
\dot{B}^{<\rho>} = \left[\sigma^{\rho}\, _{\beta} + \omega^{\rho}\,_{\beta} - \frac{2\Theta}{3} \, h^{\rho}\,_{\beta}\right]B^{\beta} - \epsilon^{\rho\mu\nu} \, \dot{u}_{\mu}E_{\nu} - \epsilon^{\rho\mu\nu} \, \nabla_{\mu}E_{\nu} \, .
\end{align}
where, $\epsilon^{\mu\nu\alpha}$ is defined in Eq.~\eqref{eq:3-LeviCivita}. The above equation provides the dynamical evolution of the magnetic field, while Eq.~\eqref{eq:Maxwell_Eq2_timelike_Final} is the constraint equation.

As mentioned above, performing simultaneous transformation $E^{\mu} \rightarrow - B^{\mu}$ and $B^{\mu} \rightarrow  E^{\mu}$ in Eqs.~\eqref{eq:Maxwell_Eq1_timelike_Final} and \eqref{eq:Maxwell_Eq1_spacelike_Final}, we obtain the time-like and space-like parts of the first Maxwell's equation~\eqref{eq:Maxwell_Eq1}:
\begin{align}
\label{eq:Maxwell_Eq1_timelike_Final}
& D_{\beta}E^{\beta} = - 2\omega^{\nu}B_{\nu} \\
\label{eq:Maxwell_Eq1_spacelike_Final}
& \dot{E}^{<\rho>} = \left[\sigma^{\rho}\, _{\beta} + \omega^{\rho}\,_{\beta} - \frac{2\Theta}{3} \, h^{\rho}\,_{\beta}\right]E^{\beta} + \epsilon^{\rho\mu\nu} \, \dot{u}_{\mu}B_{\nu} + \epsilon^{\rho\mu\nu} \, D_{\mu}B_{\nu} \, .
\end{align}
Similarly, the above equation provides the dynamical evolution of the electric field, while Eq.~\eqref{eq:Maxwell_Eq1_timelike_Final} is the constraint equation.

\subsection{In 1+1+2 formalism}

We aim to calculate the memory effect of EM fields. As the memory vector resides in the $2$-surface orthogonal to the direction of propagation of the in-coming wave, we need to decompose the 3-space to $1 + 2$-space w.r.t a given spatial direction.  
In this subsection, we rewrite Maxwell's equations (\ref{eq:Maxwell_Eq1}, \ref{eq:Maxwell_Eq2}) using the space-like vector $n^{\nu}$ and the projection tensor~\eqref{eq:projection-1+1+2} in $1+1+2$ formalism. 

To do this,  we first express the EM fields and currents in $3$-space into $1+2$ form:
\begin{align}
\label{eq:epsilon_1+2}
&E^{\mu}=\mathcal{E} n^{\mu}+\mathcal{E}^{\mu} \\
\label{eq:B_1+2}
&B^{\mu}=\mathcal{B} n^{\mu}+\mathcal{B}^{\mu} \, .
\end{align}
where, $\mathcal{E} \equiv E^{\mu}n_{\mu} $, $\mathcal{E}^{\mu} \equiv \Tilde{h}^{\mu}\,_{\nu} 
E^{\nu}$, $\mathcal{B} \equiv B^{\mu} n_{\mu}$, 
and $\mathcal{B}^{\mu} \equiv 
\Tilde{h}^{\mu}\,_{\nu} B^{\nu}$. Following the discussion in Sec.~\eqref{subsec:1+1+2}, it follows that $\epsilon_{\mu\nu}\mathcal{E}^{\nu}$ is 
orthogonal to $\mathcal{E}^{\mu}$ and, similarly,  $\epsilon_{\mu\nu}\mathcal{B}^{\nu}$ is orthogonal to $\mathcal{B}^{\mu}$. If electric and magnetic fields are orthogonal to each other in $2$ space, then we have 
\begin{align}
\label{eq:BPerpE}
\mathcal{E}^{\nu} = \epsilon_{\mu\nu}\mathcal{B}^{\nu} \quad \mathcal{B}^{\nu} = - \,\epsilon_{\mu\nu}\mathcal{E}^{\nu} \, .   
\end{align}
These relations will play an important role in Sec.~\eqref{sec:memory_generic} to derive the memory effect.

The second step is to split the evolution equations~(\ref{eq:Maxwell_Eq2_spacelike_Final}, \ref{eq:Maxwell_Eq1_spacelike_Final}) interms of $\mathcal{E}, \mathcal{E}^{\mu}, \mathcal{B}, \mathcal{B}^{\mu}$. To do that, we project Eq.~\eqref{eq:Maxwell_Eq1_spacelike_Final} along spacelike direction $n^{\mu}$ and 
multiply Eq.~\eqref{eq:Maxwell_Eq1_spacelike_Final} 
with projection tensor~\eqref{eq:projection-1+1+2}. After a long calculation, we obtain the following evolution equations for $\mathcal{E}$ (along $n^{\mu}$) and $\mathcal{E}^{\mu}$ (in the orthogonal $2$-space):
\begin{align}
\label{eq:Edot_1space_part}
 \dot{\mathcal{E}} + \Theta \mathcal{E} =&~\alpha^{\mu} \mathcal{E}_{\mu} -2 \tilde{\omega} \mathcal{B}+\epsilon_{\mu \rho} \tilde{D}^{\mu} \mathcal{B}^{\rho}
\\
\nonumber
\dot{\mathcal{E}}_{\bar{\mu}} 
+ \frac{\Theta }{2} \mathcal{E}_{\mu} 
=& - \left( \alpha_{\mu} 
+ 2 \epsilon_{\mu \rho} \Omega^{\rho} \right) \mathcal{E}
+ \left( 
\Sigma_{\mu \rho} 
+ \Omega \epsilon_{\mu \rho}
\right) \mathcal{E}^{\rho} 
+ \epsilon_{\mu \rho} \left( 
\mathcal{A}^{\rho} 
- n^{\prime \rho}  
+ \tilde{D}^{\rho} \right) \mathcal{B} 
\\
\label{eq:Edot_2space_part}
& - \epsilon_{\mu \rho} 
\left( \mathcal{A}  \mathcal{B}^{\rho} 
+ \mathcal{B}^{\prime \rho}
- \left(\tilde{D}^{\rho} \mathcal{B}_{\nu}\right) n^{\nu} \right) \, ,
\end{align}
where, $\Tilde{\omega} = \Tilde{\omega}_{\mu\nu}\, \epsilon^{\mu\nu}$,  $\Theta$ is the expansion factor defined in Eq.~\eqref{eq:cov_derivative_u_final},  $\mathcal{A}^{\mu}$ is the relative acceleration vector in 2-space defined in Eq.~\eqref{eq:udot_1+2}, $\tilde{\omega}$ is the vorticity defined in Eq.~\eqref{eq:cov_derivative_n}. $\Omega^{\mu}$, $\Omega$ is defined in Eq.~\eqref{eq:omega_1+2} and $\Sigma_{\mu\nu}$ is in Eq.~\eqref{eq:sigma_1+2}. The $2$-space component of $\dot{n}^{\mu}$ is $\alpha^{\mu}$ which is defined in Eq.~\eqref{eq:ndot_1+2}, whereas $\mathcal{A} = n^{\mu}\dot{u}_{\mu} = -u^{\mu}\dot{n}_{\mu} $ mentioned in Eq.~(\ref{eq:udot_1+2}), (\ref{eq:ndot_1+2}).  

We want to highlight the following points regarding the above expressions: First, the above equations generalize Ampere's law for arbitrary space-time. For example, in Eq.~\eqref{eq:Edot_1space_part}, the first term in the LHS corresponds to the time derivative of the electric field along space-like direction $n^{\mu}$ and the last term in RHS is the curl of the magnetic field in $2$-space. Similarly, the LHS of Eq.~\eqref{eq:Edot_2space_part} is the time derivative of the electric field in $2$-space, and in the last term in the RHS is the curl of $\mathcal{B}^{\rho}$.
Second, in the flat space-time, the expansion factor ($\Theta$), the relative acceleration vector ($\alpha^{\mu}$), and vorticity ($\Tilde{\omega}$) vanish, and the above expression lead to Ampere's law in flat space-time. Thus, background space-time introduces new couplings between the electric and magnetic field components.
Lastly, we showed that the simultaneous transformation $E^{\mu} \rightarrow - B^{\mu}$, $B^{\mu} \rightarrow  E^{\mu}$ leads to ${F^*}^{\mu\nu} \rightarrow F^{\mu\nu}$. Substituting $\mathcal{E} \rightarrow \mathcal{B} $; $\mathcal{E}^{\mu} \rightarrow \mathcal{B}^{\mu} $ and $\mathcal{B} \rightarrow - \mathcal{E} $; $\mathcal{B}^{\mu} \rightarrow - \mathcal{E}^{\mu}$ in 
Eqs.~(\ref{eq:Edot_1space_part}, \ref{eq:Edot_2space_part}), we have:
\begin{align}
\label{eq:Bdot_1space_part}
\dot{\mathcal{B}}  + \Theta \mathcal{B} =&  \mathcal{B}^{\mu} \alpha_{\mu} +2 \tilde{\omega} \mathcal{E}-\epsilon_{\mu \rho} \tilde{D}^{\mu} \mathcal{E}^{\rho} \\
\nonumber
 \dot{\mathcal{B}}_{\bar{\mu}} 
+ \frac{1}{2} \Theta \mathcal{B}_{\mu} =& 
-\left( \alpha_{\mu} + 2  \epsilon_{\mu \rho} \Omega^{\rho} \right)\mathcal{B} 
+ \left(\Sigma_{\mu \rho} +\Omega \epsilon_{\mu \rho} \right) \mathcal{B}^{\rho}
- \epsilon_{\mu \rho} \left(  \mathcal{A}^{\rho} + \tilde{D}^{\rho}  -  n^{\prime \rho}\right) \mathcal{E}\\
\label{eq:Bdot_2space_part}
& + \epsilon_{\mu \rho}\left(\mathcal{A}  \mathcal{E}^{\rho} +\epsilon_{\mu \rho} \mathcal{E}^{\prime \rho}-\left(\tilde{D}^{\rho} \mathcal{E}_{\nu}\right) n^{\nu}\right)
\end{align}
Note that we obtain the above equations by projecting Eq.~\eqref{eq:Maxwell_Eq2_spacelike_Final} along spacelike direction $n^{\mu}$ and 
multiply Eq.~\eqref{eq:Maxwell_Eq2_spacelike_Final} 
with projection tensor~\eqref{eq:projection-1+1+2}. Again, the above equations generalize Faraday's law for arbitrary space-time.

The last step is to split the constraint equations (\ref{eq:Maxwell_Eq1_timelike_Final}, \ref{eq:Maxwell_Eq2_timelike_Final}) 
interms of $\mathcal{E}, \mathcal{E}^{\mu}, \mathcal{B}, \mathcal{B}^{\mu}$. Substituting  (\ref{eq:epsilon_1+2},\, \ref{eq:B_1+2}) and the kinematic quantities (\ref{eq:udot_1+2}-\ref{eq:sigma_1+2}), we get:
\begin{align}
\label{eq:Maxwell_Eq1_timelike_Final_1+2}
\tilde{D}^{\mu} \mathcal{E}_{\mu}
+ n^{\mu} \mathcal{E}_{\mu}^{\prime}
+ \mathcal{E}^{\prime} 
+ \tilde{\Theta} \mathcal{E} 
+ 2\left(\Omega \mathcal{B}
+ \Omega^{\mu} \mathcal{B}_{\mu}\right) 
= 0 \\
\label{eq:Maxwell_Eq2_timelike_Final_1+2}
\tilde{D}^{\mu} \mathcal{B}_{\mu} 
- n^{\prime \mu} \mathcal{B}_{\mu} 
+ \mathcal{B}^{\prime} 
+ \tilde{\Theta} \mathcal{B} 
 -2\left(\Omega \mathcal{E}+\Omega^{\mu} \mathcal{E}_{\mu}\right)=0
\end{align}
where $\Tilde{\Theta}$ is the expansion along the space-like vector defined in Eq.~\eqref{eq:cov_derivative_n}.  The above equations are generalizations of Gauss law. Here again, in the flat space-time, the expansion factor ($\Tilde {\Theta}$), the relative acceleration vector ($\alpha^{\mu}$), vorticity (${\Omega}$) vanish, and the above expressions lead to Gauss law in flat space-time.

\subsection{Energy-momentum tensor of the electromagnetic field}

As we will show in the next section, the 
electromagnetic stress tensor plays a crucial role in understanding the memory effect. This subsection evaluates the electromagnetic stress tensor in $1 + 1 + 2$ formalism for an arbitrary space-time. The EM action in an arbitrary background is:
\begin{equation}
\label{eq:EMaction}
S=- \frac{1}{4} \int d^4x~\sqrt{-g}~F_{\mu \nu} F_{\rho \sigma} g^{\mu \rho} g^{\nu \sigma} \, .
\end{equation}
Varying the above action w.r.t the metric $(g^{\mu\nu})$ leads to the following energy-momentum tensor: 
\begin{equation}
T_{\mu \nu}=\frac{1}{2} g^{\rho \sigma} F_{\mu \rho} F_{\nu \sigma} -\frac{1}{8} g_{\mu \nu} g^{\rho \sigma} g^{\alpha\beta} F_{\rho \alpha} F_{\sigma \beta} \, .
\label{eq:EMStressTensor}
\end{equation}
In $1 + 3$-formalism, the stress-tensor of matter field ($T_{\mu\nu}$) can written as:
\begin{align}
\label{eq:EM-tensor-1+3}
T_{\mu\nu} =  \rho\, u_{\mu} u_{\nu} + 2\,  {S}_{(\mu} \, u_{\nu)} + {W}_{\mu\nu} \, ,
\end{align}
where, the energy-density  $\rho$, the energy flux $S^{\alpha}$
and stress-tensor $W^{\alpha\beta}$ as measured in the observer's worldline are given by~\cite{1982-Thorne.MacDonald-MNRAS}:
\begin{equation}
\rho=\mathrm{T}^{\mu \nu} 
{u}_\mu {u}_{\nu}, \quad 
{S}^\alpha=- h_\mu^\alpha \, {T}^{\mu\nu} u_\nu, \quad 
W^{\alpha \beta}= 
h_\mu^\alpha \, {T}^{\mu \nu} h_\nu^\beta  
\end{equation}
For the electromagnetic fields in $1 + 3$-formalism, $\rho,\, S_{\mu}$ and ${W}_{\mu\nu}$ are: 
\begin{align}
\label{eq:rho,poynting}
& \rho \equiv 
\frac{1}{2}\left(E^{\mu}E_{\mu} + B^{\mu}B_{\mu} \right); \quad S_{\mu } \equiv  \epsilon_{\mu\nu\rho}E^{\nu}B^{\rho}\\
\label{eq:stresstensor}
& {W}_{\mu\nu} \equiv \frac{1}{2}\left(E^{\mu}E_{\mu} + B^{\mu}B_{\mu} \right) h_{\mu\nu} - 
E_{\mu}E_{\nu} - B_{\mu}B_{\nu}
\end{align}
Rewriting $\rho$ interms of the variables ($\mathcal{E}, \mathcal{E}^{\mu}, \mathcal{B}, \mathcal{B}^{\mu})$ in $1 + 1 + 2$ formalism, we have:
\begin{align}
\label{eq:rho}
\rho &= \frac{1}{2}\left( \mathcal{E}^2+ \mathcal{B}^2\right) + \frac{1}{2}\left(\mathcal{E}^{\mu}\mathcal{E}_{\mu} + \mathcal{B}^{\mu}\mathcal{B}_{\mu} \right) = \rho_{(n)} + \rho_{\rm 2-space}
\end{align}
Thus, $\rho_{(n)}$ corresponds to the energy of the EM field along $n_{\mu}$ and $\rho_{\rm 2-space}$ corresponds to the energy of the EM field in the 2-space. The energy flux $S_{\mu}$ (a vector in $3$-space) can be rewritten in $1+2$ space as:
\begin{align}
\label{eq:Poynting_1+2}
{S}_{\mu} &= \mathcal{S} n_{\mu} + \mathcal{S}_{\mu}
\end{align}
where $\mathcal{S}$ is the Poynting vector of the EM field along the space-like vector $n^{\mu}$ and $\mathcal{S}_{\mu}$ is the energy flux in the 2-space. These are given by:
\begin{align}
\label{eq:S1}
\mathcal{S} &= {S}_{\mu}n^{\mu} = \epsilon_{\mu\nu} \mathcal{E}^{\mu}\mathcal{B}^{\nu} \, \\
\label{eq:S2}
\mathcal{S}_{\mu} &= - \epsilon_{\mu\nu}\left( \mathcal{E}\mathcal{B}^{\nu} - \mathcal{B}\mathcal{E}^{\nu}\right) = 
-\left( \mathcal{E}\mathcal{E}^{\nu} + \mathcal{B}\mathcal{B}^{\nu}\right)
\end{align}
In deriving the last expression, we have used the orthogonality condition between the electric and magnetic fields in the $2$-space, i. e., $\mathcal{E}_{\nu} = \epsilon_{\nu\mu} \mathcal{B}^{\mu}$. As we will see in the next section, the memory vector depends on the part of the electromagnetic energy density $\rho$ 
and  $\mathcal{S}_{\mu}$. 

\section{Memory effect in arbitrary space-time}
\label{sec:memory_generic}

Having written Maxwell's equations in $1 + 1 + 2$ formalism for an arbitrary space-time, we now evaluate the memory effect. Usually, in the literature, one uses the Lorentz force equation to derive EM memory. The equation of motion of a charged body (of mass $m$ and charge $e$) in both
gravitational and electromagnetic fields are:
\begin{equation}
\label{eq:LorenzForce}
 m \frac{d u_\alpha}{d \tau} - 
 \frac{m}{2} g_{\beta \gamma, \alpha} u^\beta u^\gamma =e F_{\alpha \beta} u^\beta   
\end{equation}
However, the above expression does not consider the new couplings between the electric and magnetic field components in Eqs. \eqref{eq:Edot_1space_part} - \eqref{eq:Bdot_2space_part}. Hence, we use the complete Maxwell's equations \eqref{eq:Edot_1space_part} - \eqref{eq:Maxwell_Eq2_timelike_Final_1+2} and explicitly obtain the change in velocity ($\Delta u^{\mu}$) of the time-like observer. More specifically, using Eqs.~(\ref{eq:Edot_2space_part}, \ref{eq:Bdot_2space_part}), we first calculate the acceleration vector $\mathcal{A}^{\mu}$ in the $2$-space. We can then integrate the expression for the acceleration vector ($\mathcal{A}^{\mu}$ in the $2$-space) with respect to time $t$ or null time coordinate $u \equiv (t - r)$ leading to the memory vector. 

In the rest of this section, we calculate $\mathcal{A}^{\mu}$ for observers whose 
tangents are congruent to the space-like geodesics. This implies $n^{\sigma}D_{\sigma} n^{\rho} = n^{\prime \rho} = 0$, i. e., $n^{\mu}$ is tangent to a congruence of space-like geodesics~\cite{2021-Mavrogiannis.Tsagas-PRD}. Using this condition and 
substituting $\dot{\mathcal{E}}_{\Bar{\mu}} =\Tilde{h}_{\mu\nu}\dot{\mathcal{E}}^{\nu},\, \mathcal{B}^{\prime\, \rho}= n^{\nu}D_{\nu}\mathcal{B}^{\rho} $ in  Eqs.~(\ref{eq:Edot_2space_part}, 
\ref{eq:Bdot_2space_part}), we get:

%
\begin{align}
\nonumber
\Tilde{h}_{\mu\nu}\dot{\mathcal{E}}^{\nu} + \epsilon_{\mu \rho} n^{\nu}D_{\nu}\mathcal{B}^{\rho} =& 
-\frac{1}{2} \Theta \mathcal{E}_{\mu}  - \left(\alpha_{\mu} + 2  \epsilon_{\mu \rho} \Omega^{\rho}\right)\mathcal{E}
+ \left(\Sigma_{\mu \rho} +\Omega \epsilon_{\mu \rho} \right)\mathcal{E}^{\rho} \\
\label{eq:Edot_2space_part_simplified} 
& +\left(\epsilon_{\mu \rho} \mathcal{A}^{\rho} 
 + \epsilon_{\mu \nu} \tilde{D}^{\nu}   \right)\mathcal{B} 
-\epsilon_{\mu \nu}\left(\tilde{D}^{\nu} n^{\rho}\right) \mathcal{B}^{\rho} - \epsilon_{\mu \rho} 
 \mathcal{A}  \mathcal{B}_{\rho}\\
\nonumber
\left(\Tilde{h}_{\mu\nu}\dot{\mathcal{B}}^{\nu} - \epsilon_{\mu \rho} n^{\nu}D_{\nu}\mathcal{E}^{\rho}\right) &= -\frac{1}{2} \Theta  \mathcal{B}_{\mu} -   \left( \alpha_{\mu} + 2  \epsilon_{\mu \rho} \Omega^{\rho}\right)\mathcal{B}
+ \left( \Sigma_{\mu \rho}  +  \Omega \epsilon_{\mu \rho} 
\right)\mathcal{B}^{\rho}
\\
\label{eq:Bdot_2space_part_simplified} 
& - \left( \epsilon_{\mu \rho} \mathcal{A}^{\rho} 
+\epsilon_{\mu \nu} \tilde{D}^{\nu} \right)\mathcal{E}+ \epsilon_{\mu \nu} \left(\tilde{D}^{\nu} n^{\rho} \right) \mathcal{E}_{\rho}  + \epsilon_{\mu \rho} 
 \mathcal{A}  \mathcal{E}^{\rho}
\end{align}
Multiplying Eq.~\eqref{eq:Edot_2space_part_simplified} with $\mathcal{B}$, multiplying Eq.~\eqref{eq:Bdot_2space_part_simplified} with $\mathcal{E}$ and subtracting the resultant equations leads to: 
\begin{align}
\nonumber
 \epsilon_{\mu\nu} \mathcal{A}^{\nu} & = - \, \frac{\epsilon_{\mu\nu}}{2} \, \frac{D^{\nu}(\mathcal{E}^2 + \mathcal{B}^2)}{(\mathcal{E}^2 + \mathcal{B}^2)}  
+ \left(\Sigma_{\mu\nu} + \Omega \epsilon_{\mu\nu} - \frac{\Theta}{2} \Tilde{h}_{\mu\nu}\right)  \, 
\frac{(\mathcal{E} \mathcal{B}^{\nu} - \mathcal{B} \mathcal{E}^{\nu})}{(\mathcal{E}^2 + \mathcal{B}^2)} \\
 \nonumber
&+ \epsilon_{\mu\nu} \, 
\left(\Tilde{\sigma}^{\rho\nu} + \Tilde{\omega}^{\rho\nu} + 
\frac{\Tilde{\Theta}}{2} \Tilde{h}^{\rho\nu}
\right) \, 
\frac{ (\mathcal{B}\mathcal{B}_{\rho}+\mathcal{E}\mathcal{E}_{\rho})}{(\mathcal{E}^2 + \mathcal{B}^2)}  + \frac{\epsilon_{\mu \rho} 
 \mathcal{A} \left( \mathcal{E}\mathcal{E}^{\rho}  +  \mathcal{B} \mathcal{B}^{\rho}\right)}{(\mathcal{E}^2 + \mathcal{B}^2)}\\
\label{eq:memory_generic_w/o_orthogonality}
& + \frac{\mathcal{B}}{(\mathcal{E}^2 + \mathcal{B}^2)} \left(\Tilde{h}_{\mu\nu}\dot{\mathcal{E}}^{\nu} + \epsilon_{\mu \rho} n^{\nu}D_{\nu}\mathcal{B}^{\rho}\right) 
- \frac{\mathcal{E}}{(\mathcal{E}^2 + \mathcal{B}^2)}   \left(\Tilde{h}_{\mu\nu}\dot{\mathcal{B}}^{\nu} - \epsilon_{\mu \rho} n^{\nu}D_{\nu}\mathcal{E}^{\rho}\right)
\end{align}
To have a transparent understanding, we substitute the definitions \eqref{eq:rho} - \eqref{eq:S2} in the  expression above, resulting in:
\begin{align}
\nonumber
 \epsilon_{\mu\nu} \mathcal{A}^{\nu} & = - \, \frac{\epsilon_{\mu\nu}}{2} \, \frac{D^{\nu}\rho_{(n)}}{\rho_{(n)}}  
- \frac{\epsilon^{\nu\alpha}}{2} \left(\Sigma_{\mu\nu} + \Omega \epsilon_{\mu\nu} - \frac{\Theta}{2} \Tilde{h}_{\mu\nu}\right)  \, 
\frac{\mathcal{S}_{\alpha}}{\rho_{(n)}} 
- \frac{\epsilon_{\mu\nu}}{2} \, 
\left(\Tilde{\sigma}^{\rho\nu} + \Tilde{\omega}^{\rho\nu} + 
\frac{\Tilde{\Theta}}{2} \Tilde{h}^{\rho\nu}
\right) \, 
\frac{ \mathcal{S}_{\rho}}{\rho_{(n)}} \\
\label{eq:memory_generic}
&\,  - \frac{\epsilon_{\mu \rho} 
  \mathcal{S}^{\rho}\, \mathcal{A}}{2 \rho_{(n)}} + \frac{\mathcal{B}}{2\rho_{(n)}} \left(\Tilde{h}_{\mu\nu}\dot{\mathcal{E}}^{\nu} + \epsilon_{\mu \rho} n^{\nu}D_{\nu}\mathcal{B}^{\rho}\right) 
- \frac{\mathcal{E}}{2\rho_{(n)}}   \left(\Tilde{h}_{\mu\nu}\dot{\mathcal{B}}^{\nu} - \epsilon_{\mu \rho} n^{\nu}D_{\nu}\mathcal{E}^{\rho}\right) \,.
\end{align}
This is the master equation for the EM memory in arbitrary space-time regarding which we would like to discuss the following points: First, to our understanding, this is a first time the EM memory has been obtained for 
an arbitrary space-time. In the previous calculations~\cite{Bieri:2013hqa,Winicour:2014ska}, the authors have restricted to asymptotic flat space-times. 
Second, the last two terms in the RHS of the above expression vanishes in the asymptotic limit. To see this, let us consider a spherically symmetric space-time. Let $t$ refer to the time coordinate and $r$ to the radial coordinate and the null coordinate is $u \equiv t -r$. In the asymptotic limit $\partial_{u} \sim \partial_{t}$  and $\partial_{u} \sim -\partial_{r}$. Setting $u^{\mu} \equiv (1,\, 0,\, 0,\, 0)$ and $n^{\mu} \equiv (0,\, 1,\, 0,\, 0)$, 
the penultimate term in the RHS of the above equation simplifies to:
\begin{align}
\Tilde{h}_{\mu\nu}\dot{\mathcal{E}}^{\nu} + \epsilon_{\mu \rho} n^{\nu}D_{\nu}\mathcal{B}^{\rho} &\simeq \Tilde{h}_{\mu\nu}u^{0} \nabla_{0}\mathcal{E}^{\nu} + \epsilon_{\mu \rho} n^{1}\nabla_{1}\mathcal{B}^{\rho} \simeq \Tilde{h}_{\mu\nu}\partial_{u} \mathcal{E}^{\nu} - \epsilon_{\mu \rho} \partial_{u}\mathcal{B}^{\rho} \nonumber \\
\label{eq:E_B_asymptotic}
&= f(u)\partial_{u}\left(\Bar{\Tilde{h}}_{\mu\nu} \mathcal{E}^{\nu} - \Bar{\mathcal{E}}_{\mu\nu} \mathcal{B}^{\nu}\right)
\end{align}
where, $\Tilde{h}_{\mu\nu}=f(u)\Bar{\Tilde{h}}_{\mu\nu}$ and $\Bar{\epsilon}_{\mu\nu}= f(u)\epsilon_{\mu\nu}$. The terms with bar represent their time independent parts. 
The above expression vanishes if $\mathcal{E}^{\nu}$ and $\mathcal{B}^{\nu}$ are orthogonal to each other in the 2-space. As we mentioned earlier \eqref{eq:BPerpE},
in 2-space, the electric and magnetic fields are always orthogonal to each other. Similarly, the last term can also be shown to vanish in the asymptotic limit.
Thus, the above master equation \eqref{eq:memory_generic} reduces to:
\begin{align}
\nonumber
 \epsilon_{\mu\nu} \mathcal{A}^{\nu} = & - \, \frac{\epsilon_{\mu\nu}}{2} 
\frac{D^{\nu}\rho_{(n)}}{\rho_{(n)}}  
- \frac{\epsilon^{\nu\alpha}}{2} \left(\Sigma_{\mu\nu} + \Omega \epsilon_{\mu\nu} - \frac{\Theta}{2} \Tilde{h}_{\mu\nu}\right)  
\frac{\mathcal{S}_{\alpha}}{\rho_{(n)}} \\
& \, - \frac{\epsilon_{\mu\nu}}{2} 
\left(\Tilde{\sigma}^{\rho\nu} + \Tilde{\omega}^{\rho\nu} + 
\frac{\Tilde{\Theta}}{2} \Tilde{h}^{\rho\nu}
\right) \frac{ \mathcal{S}_{\rho}}{\rho_{(n)}} - \frac{\epsilon_{\mu \rho} 
 }{2 \rho_{(n)}}\,  \mathcal{S}^{\rho}\, \mathcal{A}
\label{eq:memory_generic_w/o_orthogonality_final}
\end{align}
Third, the above expression provides a nice geometrical understanding of the various contributions to memory effect. The first term in the RHS corresponds to the change in the EM field energy ($\rho_{(n)}$) along $n_{\mu}$
in the 2-space. This does not contain any contribution from the kinematical properties of the space-time. In other words, this term will vanish if the EM field energy does not change in the 2-space, like a 2-D flat sheet. However, as we show in the next section, this is non-zero in  flat space-time expressed in spherical coordinates. The next two terms in the RHS are proportional to the energy flux ($\mathcal{S}_{\alpha}$) in the 2-space. However, both these terms have different kinematical information of the space-time and vanish for flat space-time. The second term in the RHS carries  information about shear $(\Sigma_{\mu\nu})$, 
vorticity scalar $(\Omega)$ related to $n^{\mu}$ and expansion scalar $(\Theta)$ corresponding to time-like observer $u^{\mu}$. The third term in the RHS carries information about shear $(\Tilde{\sigma}^{\mu\nu})$, 
vorticity tensor $(\Tilde{\omega}^{\mu\nu})$ and expansion scalar $(\Tilde{\Theta})$ corresponding to the space-like vector $n^{\mu}$.

Fourth, as mentioned earlier, we have not included external currents or charges in our analysis. Hence, the acceleration vector does not have contribution from the external sources. Hence, the memory vector we obtain is equivalent to the null-kick derived in Refs.~\cite{Bieri:2013hqa,Winicour:2014ska}. 
It is also important to note that these authors did not obtain the contributions due to the kinematical properties of the space-time. However, as we will see in the next section, their contribution can be significant. 

Lastly, to obtain the memory vector, we need to integrate the above expression w.r.t the proper time of the observer --- $\Delta u^{\mu}$ is the memory vector. It is interesting to note that initially if the observer has non-zero velocity \emph{only} along the time direction, at a later time, 
due to the memory effect, there is a non-zero velocity in the 2-space.



\section{Application to specific space-times}
\label{sec:memory_specific_sptm}

In the previous section, we obtained a master equation \eqref{eq:memory_generic} for the EM vector for an arbitrary 4-D space-time using $1 + 1+ 2$-formalism.

As discussed in the previous section, the master equation \eqref{eq:memory_generic} corresponding to the acceleration of the comoving observer in the 2-D surface provides a physical understanding of the contribution to the memory. For instance, the leading order contribution only requires information about the total energy density of the EM field, while the subleading contributions contain information about the space-time geometry and the other components of the energy-momentum tensor of the EM field. This is the first time a transparent and easily applicable final expression for electromagnetic memory has been derived for general curved space-time. In order to illustrate this fact, we consider 
specific examples and obtain the memory vector. In this section we obtain memory vector for flat,  FLRW, \emph{pp-wave}, and Kerr space-times.  

\subsection{Minkowski space-time}
\label{subsec:memory_Minkowski}
In order to compare the master equation with the existing results~\cite{Bieri:2013hqa}, we first consider Minkowski space-time in spherical coordinates: 
\begin{align}
   \label{eq:metric_Minkowski}
   ds^{2} = - dt^2 + dr^2 + r^{2}\, \gamma_{A B} 
\end{align}
where, 
\begin{align}
\gamma_{A B} = \left(\begin{array}{cc}
1  & 0  \\
 0 & \sin^{2}\theta
\end{array}\right)
\end{align}
is the metric describing unit $2$-sphere. In Minkowski space-time, the $4$-velocity of the time-like congruence observer is $u^{\mu} \equiv (1,\, 0,\, 0,\, 0)$ and the space-like vector is $n^{\mu}  \equiv (0,\, 1,\, 0,\, 0) $. Since $\nabla_{\mu}u_{\nu} = 0$ and  $\nabla_{\mu}n_{\nu} = 0$, the \textit{kinematics} quantities, defined in Sec.~(\ref{subsec:1+3},  \ref{subsec:1+1+2}) vanish for the Minkowski space-time. Hence only the first term in Eq.~\eqref{eq:memory_generic} will be non-zero, i. e.,
\begin{align}
\label{eq:memory_minkowski}
\mathcal{A}_{\rm Flat}^{\nu} & = - \, \frac{1}{2} \, \frac{D^{\nu} \rho_{\rm n} }{\rho_{\rm n}} \, .
\end{align}
As mentioned earlier, the acceleration vector corresponds to acceleration in the 2-Sphere. 
Hence, it is appropriate to switch to the 2-Sphere index: 
\[
\mathcal{A}^{A} = u^{\mu}\nabla_{\mu} u^{A} = u^{0}\partial_{0} u^{A} \, + \, 2 u^{0} \Gamma^{A}_{0\, B} u^{B} \, . 
\]
Since the 4-velocity $u^{\mu}$ is zero in the 
2-Sphere, we have $\mathcal{A}^{A} = u^{0}\partial_{0} u^{A} = \partial_{t}u^{A}$. In null coordinate, this becomes $\mathcal{A}^{A} =\partial_{t}u^{A}$. Substituting the above expression in Eq.~\eqref{eq:memory_minkowski} and integrating in the null coordinate, we have:
%
\begin{align}
\label{eq:memory_Minkowski1}
\Delta u^{A} \equiv \int\, du \, \mathcal{A}^{A}  = -\frac{1}{2}\, \int\, du  \, \frac{D^{A} \rho_{\rm n} }{\rho_{\rm n}} \, .
\end{align}
The above expression is velocity kick w.r.t the  \emph{Lagrangian observers}. To compare this with the net momentum (kick) vector as seen by the asymptotic static observers (\emph{Eulerian observers}), we need to do a coordinate transformation. 
Specifically, we need to transform from coordinate basis $\left(\Vec{e}^{\, \theta}, \Vec{e}^{\, \phi}\right)$ to orthogonal coordinate basis $\left(\hat{\theta}, \hat{\phi}\right)$.
In terms of 
$\left(\hat{\theta}, \hat{\phi}\right)$, we have
$\Delta \Vec{u} \equiv \Delta u^{\mu} \Vec{e}_{\mu}$, where, $\Vec{e}^{\, \theta} = \hat{\theta}/r \, , 
\, \Vec{e}^{\, \phi} = \hat{\phi}/(r\, \sin \theta)$. Thus, the velocity kick w.r.t the  asymptotic static observers is given by:
\begin{align}
\label{eq:memoryvector_Minkowski}    
\Delta \Vec{u}_{\rm Flat} & = \frac{1}{r} 
\left( \Delta u^{\theta} \hat{\theta} +
\frac{\Delta u^{\phi}}{\sin \theta} \hat{\phi} \right) 
\end{align}
Interestingly, the EM memory vector in Minkowski space-time is inversely proportional to ${r}$ and matches with Ref.~\cite{Bieri:2013hqa}. 
This passes the first test that the master equation \eqref{eq:memory_generic} indeed describes the EM memory vector for a static asymptotic observer. In the rest of this section, we obtain the memory vector for non-flat geometries and show the robustness of our approach.

\subsection{FLRW space-time}
\label{subsec:memory_FLRW}
The conformally flat FLRW metric in spherical coordinates is:
\begin{align}
\label{eq:metric_FLRW}
ds^2 = a(\eta)^2 \, \left(- d\eta^2 + dr^2 + r^{2}\, \gamma_{A B} \right)
\end{align}
where, the conformal time ($\eta$) is related to the cosmic time ($t$) by $dt = a(\eta)\, d\eta$. In $1+3$ formalism, the fundamental observer with time-like $4-$velocity in FLRW metric is $u^{\mu} = dx^{\mu}/dt =  dx^{\mu}/( a(\eta))\, d\eta = \left(\,1,\,0,\,0,\,0\,\right)/a(\eta)$. For this choice of observer, the $3-$space projection tensor $(h_{\mu \nu})$ orthogonal to $u^{\mu}$ is:
\begin{align}
\label{eq:3-metric-FLRW}
h_{\mu \nu}=\left(\begin{array}{cc}
 a^2(\eta)  & 0  \\
 0 & a^2(\eta) \, r^{2}\, \gamma_{A B}
\end{array}\right) \, .
\end{align}
Since the FLRW line-element is homogeneous and isotropic,  only the expansion scalar ($\Theta$) is non-zero: 
\[
\Theta = 3 \frac{{\cal H}(\eta)}{a(\eta)} 
\quad \mbox{where} 
\quad {\cal H} = \frac{a'(\eta)}{a(\eta)} 
\]
where $'$ refers to derivative w.r.t. $\eta$. Other kinematic quantities vanish, i. e., $\sigma_{\mu\nu} = \omega_{\mu\nu} = 0$. 

We now spilt the 3-space into $1+2$ by choosing the following space-like vector $n^{\mu} = 
(0,\,{1},\,0,\,0)/{a(\eta)}$. This satisfies the conditions: $n^{\mu}n_{\mu} = 1$ and $u^{\mu}n_{\mu} = 0$. Repeating the steps discussed in Sec.~(\ref{subsec:1+1+2}) for the line-element \eqref{eq:metric_FLRW}, we get: 
\[
\Tilde{\Theta} = \frac{2}{a(\eta)} \frac{1}{r} \, , 
\Tilde{\sigma}_{\mu\nu} = \Tilde{\omega}_{\mu\nu}=0 .
\]
It is important to note that while $\Theta$ is a function of $\eta$ only, 
$\Tilde{\Theta}$ depends on both $\eta$ and $r$. 
Also, $\Theta$ depends on the Hubble parameter ${\cal H}$, while $\Tilde{\Theta}$ is inversely proportional of $r$. Hence, at large distances, $\Tilde{\Theta}$ decays faster compared to $\Theta$. 

Substituting the above expressions in Eq.~\eqref{eq:memory_generic_w/o_orthogonality_final}, we have:
\begin{align}
\label{eq:memory_FLRW}
\mathcal{A}^{\nu}_{\rm FLRW} & = - \, \frac{1}{2} \, \frac{D^{\nu} \rho_{\rm n} }{\rho_{\rm n}}+ \frac{1}{4\, \rho_{\rm n}} \, \mathcal{S}^{\nu}(\Theta -\Tilde{\Theta}) \, .
\end{align}

Like Minkowski space-time, $\mathcal{A}^{\nu}$ will have components only in 
the $2$-Sphere. Using the fact that the fundamental observers have zero velocity in the 2-Sphere and 
repeating the earlier analysis, we have
\[
\mathcal{A}^{A} = u^{0}\partial_{0} u^{A} 
= \frac{1}{a(\eta)} \, 
\frac{\partial u^{A}}{\partial \eta} \, .
\]
In terms of the null coordinate $u (\equiv \eta - r)$, we have:
\[
\mathcal{A}^{A}  = \frac{1}{a(u)} 
\, \frac{\partial u^{A}}{\partial u} \, .
\]
Substituting the above expression in Eq.~\eqref{eq:memory_FLRW}, we have:
\begin{align}
\label{eq:memory_FLRW_null_coor}
\frac{\partial u^{A}}{\partial u}   = -  \, \frac{a(u)}{2} \, \frac{D^{A} \rho_{\rm n} }{\rho_{\rm n}}+ \frac{a(u)}{4\, \rho_{\rm n}} \, \mathcal{S}^{A}(\Theta -\Tilde{\Theta}) \, .
\end{align}
Integrating the above expression w.r.t $u$, leads to the following memory vector:
\begin{align}
\label{eq:memory_vector_FLRW_null_coor} 
\Delta u_{\rm FLRW}^{A} =   - \frac{1}{2} \, \int du \, \frac{a(u)}{\rho_{\rm n}} \, D^{A} \rho_{\rm n}  + \frac{1}{4} \, \int du \, \frac{a(u)}{\rho_{\rm n}} \, \mathcal{S}^{A}(\Theta -\Tilde{\Theta})
\end{align}
This is the expression for the memory vector in FLRW space-time regarding which we want to highlight the following points: First, unlike Minkowski space-time, here the fundamental observers are Lagrangian, and hence, we do not have to transform the above expression to Lagrangian observers. Second, our results differ from the results of Ref.~\cite{2022-EnriquezRojo-JHEP}. 
In Ref.~\cite{2022-EnriquezRojo-JHEP}, the authors show that the EM memory effect in FLRW differs from the Minkowski only by the conformal factor $a(\eta)$ or $a(u)$. In other words, their analysis did not account for the geometric contribution to the memory effect. As mentioned earlier, the geometric contribution leads to a non-zero energy flux ($\mathcal{S}^{A}$) contribution. Also note that the ordinary memory derived in Ref.~\cite{2022-EnriquezRojo-JHEP} is not present in Eq.~\eqref{eq:memory_vector_FLRW_null_coor}  as we have assumed any external charge or current to be zero.
Third, we find that $\rho_{(n)}$ 
and the energy flux ($\mathcal{S}^{A}$) contribute oppositely. It will be interesting to see whether the two contributions nullify the EM memory.   


\subsection{\textit{pp-wave} space-times}
\label{subsec:memory_pp-wave}

In this subsection, we derive the EM memory for 
a special kind of {plane-fronted wave with parallel rays} (pp-waves) called plane-wave metric~\cite{2003-Blau-CQG}:
\begin{align}
\label{eq:ppwavemetric}
ds^{2} = - 2 du dv - \mathcal{F}(u,\, x, \, y) \, du^{2} + dx^{ 2} + dy^{ 2} 
\end{align}
where, $\mathcal{F}(u,\, x, \, y) = A(u)(x^2 -y^2) + 2B(u)xy$ describes the plane wave and $A(u), B(u)$ are arbitrary functions such that $\mathcal{F} > 0$. Note that $u, v$ are not light-cone coordinates. $u$ is time-like coordinate and $v$ is a null coordinate.

We split the above 4-D space-time into $1+3$ form and later into $1+1+2$-form by considering the following 
time-like velocity vector ($u^{\mu}$) and space-like vector $(n^{\mu})$:
\[ 
u^{\mu}\equiv \left(\mathcal{F}(u, \, x, \,y)^{(-1/2)},\, 0,\, 0,\, 0\right), \quad 
n^{\mu} \equiv \left(\mathcal{F}(u, \, x, \,y)^{(-1/2)},\, -\, \mathcal{F}(u, \, x, \,y)^{(1/2)},\, 0,\, 0\right) \, .
\]
For the above choice of time-like vector, the 3-space projection tensor $(h_{\mu\nu})$ is:
\begin{align}
    \label{eq:ppwavemetrich}
    h_{\mu\nu} &= 
  \begin{bmatrix}
    \, 0 & 0 & 0 & 0 \, \\
    \, 0 & \frac{1}{\mathcal{F}(u,\, x,\, y)} & 0 & 0 \, \\
    \, 0 & 0 & 1 & 0 \, \\
    \, 0 & 0 & 0 & 1 \,
  \end{bmatrix}
\end{align}
Substituting these in the definitions in Sec. \eqref{sec:covariant_formalism}, only non-zero quantity is the expansion scalar ($\Theta$):
\begin{align}
\label{eq:theta_pp}
\Theta = -\frac{\left(x^2 - y^2\right) \,  A^{\prime}(u) \, + \, 2 x y \, B^{\prime}(u)}{2 \, \left(2 B(u) \, x y \, + \, A(u) \, \left(x^2-y^2\right) \, \right)^{3 / 2}} \, .
\end{align}
The non-zero projection tensor $\Tilde{h}_{\mu \nu}$ components in the 2-space are
$\Tilde{h}_{x x} = 1, \, \Tilde{h}_{y y} = 1$.
Thus, the memory vector for the special kind of pp-wave space-times is: 
\begin{align}
\label{eq:memory_pp}
\mathcal{A}_{\rm PP}^{\nu} & = - \, \frac{1}{2} \, \frac{D^{\nu} \rho_{\rm n} }{\rho_{\rm n}}+ \frac{\Theta}{4\, \rho_{\rm n}} \, \mathcal{S}^{\nu} 
\, .
\end{align}
Here, the acceleration of the time-like observer is confined to the $x-y$ plane, i. e.,
\begin{align}
\label{eq:memory_pp_2space}
\mathcal{A}_{\rm PP}^{A} & = - \, \frac{1}{2} \, \frac{D^{A} \rho_{\rm n} }{\rho_{\rm n}}+ \frac{\Theta}{4\, \rho_{\rm n}} \, \mathcal{S}^{A} \, , 
\end{align}
where, the index $A,\, B$ corresponds to $(x,\, y)$. Evaluating the acceleration vector along $x$ and $y$, we have: 
\begin{align}
\label{eq:memory_pp_2space1}
\mathcal{A}^{\rm (PP)}_{x (y)} & = - 
\frac{1}{2 \, \rho_{\rm n}} \, \partial_{x (y)} 
\left( \rho_{\rm n} \right) + \frac{\Theta}{4\, \rho_{\rm n}} \, \mathcal{S}_{x (y)} \, .
\end{align}
Integrating the above equation w.r.t $u$, we have:
\begin{align}
\label{eq:memory_vector_pp}    
\Delta u^{\rm PP}_{x( y)} = - \frac{1}{2 }\int du \,  \, \frac{\partial_{x( y)} \left(\rho_{\rm n}\right)}{\rho_{\rm n}} \, + \, \frac{\Theta}{4}\, \int du \frac{\mathcal{S}_{x( y)} }{\rho_{\rm n}} \,.  
\end{align}
%

The above expression for the velocity kick is 
for a generic plane-wave metric. To gain some physical intuition, we consider two specific forms ---  Penrose limit of the Schwarzschild and FLRW space-times~\cite{2003-Blau-CQG}. For Schwarzschild space-time, we have
\[
A(u) = \frac{6}{25 u^{2}};  \quad B(u) = 0
\]
Substituting these in Eq.~\eqref{eq:theta_pp}, we have: 
\[
\Theta_{\rm PP, Sch} = \frac{5}{\sqrt{6(x^2-y^2)}} \, .
\]
It is interesting to note that although the space-time metric does not differentiate between the two spatial coordinates $(x,y)$, in order for $\Theta$ to be real, the above expression demands that $x > y$. Thus, velocity kick due to EM wave in PP-wave limit of Schwarzschild space-time can only occur if $x > y$ and is given by:
\begin{align}
\label{eq:memory_vector_pp_Sch}    
\Delta u^{\rm PP\, Sch}_{x( y)} = - \frac{1}{2 }\int du \,  \, \frac{\partial_{x( y)} \left(\rho_{\rm n}\right)}{\rho_{\rm n}} \, + \, \frac{5}{4\sqrt{6(x^2-y^2)}}\, \int du \frac{\mathcal{S}_{x( y)} }{\rho_{\rm n}} \,.  
\end{align}
In the case of Penrose limit of FLRW space-time with power-law scale factor $a(t) \sim t^h$, we have: 
\[
A(u) = - \frac{h}{(1+h) u^{2}}, \quad B(u) = 0 \, .
\]
Substituting these in Eq.~\eqref{eq:theta_pp}, we have: 
\begin{align}
\label{eq:memory_vector_pp_FLRW}    
\Theta_{\rm PP, FLRW} = 
\sqrt{\frac{(1+h)}{h (y^2-x^2)}}; \quad 
\Delta u^{\rm PP\, FLRW}_{x( y)} = -\frac{1}{2 }\int du \, \frac{\partial_{x( y)} \left(\rho_{\rm n}\right)}{\rho_{\rm n}} +  \frac{\sqrt{(1+h)}}{4\sqrt{h(y^2-x^2)}} \int du \frac{\mathcal{S}_{x( y)} }{\rho_{\rm n}} \,. 
\end{align}
Here again, we see that in-order for $\Theta$ to be real, the above expression demands that $y > x$. Thus, velocity kick due to EM wave in PP-wave limit of FLRW space-time occurs in a different region of the 2-space compared to Schwarzschild. Thus, EM memory has a distinct signature for different space-times and can potentially be used as a probe.

\subsection{Kerr space-time}
\label{subsec:memory_kerr}

In this section, we derive the memory effect in Kerr space-time. In Boyer-Lindquist coordinates 
$\left(t, \, r, \, 
\chi , \, \phi\right)$, the Kerr space-time is:
\begin{align}
\label{eq:metric_Kerr}
d s^2= & \left(\frac{2 m r}{r^2+a^2 \chi^2}-1\right) d t^2 + \left(\frac{r^2+a^2 \chi^2}{r^2-2 m r+a^2}\right) d r^2+\left(\frac{r^2+a^2 \chi^2}{1-\chi^2}\right) d \chi^2 \nonumber \\
& - \left[\frac{4 m a r\left(1-\chi^2\right)}{r^2+a^2 \chi^2}\right] d t d \varphi + \left(1-\chi^2\right)\left[r^2+a^2+\frac{2 m a^2 r\left(1-\chi^2\right)}{r^2+a^2 \chi^2}\right] d \varphi^2 \, .
\end{align}
where $\chi \equiv \cos \theta$. In this case, 
the time-like observer 4-velocity $(u^\mu)$ 
and the space-like vector $(n^{\mu})$ 
are~\cite{2021-Hansraj.SunilMaharaj-gr-qc}:
\[
u^\mu=\left[\sqrt{\frac{r^2-2 m r+a^2}{r^2+a^2 \chi^2}}, 0, 0, 0\right],  n^{\mu} =\left[0, \sqrt{\frac{r^2-2 m r+a^2}{r^2+a^2 \chi^2}}, 0, 0\right] \, .
\]
We give below the 
kinematical quantities (discussed in Sec.~\eqref{subsec:1+1+2}) for Kerr space-time in $1+1+2$ formalism obtained in Ref.~\cite{2021-Hansraj.SunilMaharaj-gr-qc}:
\begin{align}
\label{eq:Kerrdef1}
\Theta = 0; & \quad \Sigma_{\mu\nu} = 0 \, ; \\
\Omega =  -\frac{2m a r \chi\sqrt{\mathcal{L}}}{J\sqrt{\mathcal{K}^3}} ; 
& \quad \Tilde{\Theta} = \frac{\mathcal{W}}{\mathcal{J}\sqrt{\mathcal{K}^3 \mathcal{L}}} \, ; \\
\Tilde{\omega}_{\mu\nu} = \Tilde{\omega}\epsilon_{\mu\nu} = 0 ;
& \quad \mathcal{A} = -\frac{m \mathcal{D }\sqrt{\mathcal{L}}}{J\sqrt{\mathcal{K}^3}} \, ; \\
&\Tilde{\sigma}_{\mu\nu} = \left[\begin{array}{cccc}
0 & 0 & 0 & 0 \\
0 & 0 & 0 & 0 \\
0 & 0 & -\frac{1}{2} \frac{a^2(m-r) \sqrt{\mathcal{K}}}{\mathcal{J} \sqrt{\mathcal{L}}} & 0 \\
0 & 0 & 0 & \frac{1}{2} \frac{a^2(m-r) \mathcal{M}^2 \sqrt{\mathcal{L} \mathcal{K}}}{\mathcal{J}^2}
\end{array}\right]
\end{align}
where,
\begin{align}
\mathcal{M} = \chi^{2} -1; 
& \quad \mathcal{D}=-r^2+a^2 \chi^2; 
\quad \mathcal{L}=r^2-2 m r+a^2  \\
\mathcal{J}=r^2-2 m r+a^2 \chi^2;
& \quad  \mathcal{K}=r^2+a^2 \chi^2 
\end{align}
\vspace*{-1.9cm}

\begin{align}
\label{eq:KerrdefN}
\mathcal{W} =  2 r^3(r-2 m)^2+a^4 \chi^2\left(m+r-m \chi^2+r \chi^2\right) 
 +a^2 r^2\left(-3 m+r+\chi^2(3 r-5 m)\right)
\end{align}

Substituting these expressions in Eq. \eqref{eq:memory_generic_w/o_orthogonality_final}, and noting that the memory vector lies in the 2-D surface, we get: 
\begin{align}
\label{eq:memory_vec_kerr_2space}
\mathcal{A}^{A} = & - \, \frac{1}{2} 
\frac{D^{A}\rho_{(n)}}{\rho_{(n)}}  
- \frac{\Omega}{2}   
\frac{\epsilon^{A B}\,\mathcal{S}_{B}}{\rho_{(n)}} 
\, - \frac{1}{2} 
\left(\Tilde{\sigma}^{A B}  + 
\frac{\Tilde{\Theta}}{2} \Tilde{h}^{A B}
\right) \frac{ \mathcal{S}_{B}}{\rho_{(n)}} - \frac{\mathcal{A}
 }{2 \rho_{(n)}}\,  \mathcal{S}^{A}\, 
\end{align}
This is the EM memory vector for a Lagrangian observer in Kerr space-time. Note that this is a generic result for any value of angular momentum. For a better physical insight, we consider $a \rightarrow 0$ limit. Substituting $a \to 0$ in Eqs. (\ref{eq:Kerrdef1} - \ref{eq:KerrdefN}), we have
\begin{align}
\mathcal{M}_{0} = \chi^{2} -1; 
& \quad \mathcal{D}_{0}=-r^2 ; 
\quad \mathcal{L}_{0}=r^2-2 m r  \\
\mathcal{J}_{0} =r^2-2 m r ;
& \quad  \mathcal{K}_{0} =r^2 ; \quad \mathcal{W}_{0} =  2 r^3(r-2 m)^2 \\
\Omega_{0} = \Tilde{\sigma_{0}}^{\mu\nu} = 0; 
& \quad 
\Tilde{\Theta}_{0} = 2 \sqrt{\frac{(r-2m)}{r^3}}; \quad \mathcal{A} = \frac{m}{\sqrt{r^3(r-2m)}} 
\end{align}
Substituting the above quantities in  Eq.~\eqref{eq:memory_vec_kerr_2space}, we have:
\begin{align}
\label{eq:memory_vec_kerr_2space_sim}
\mathcal{A}^{A} =  - \, \frac{1}{2} 
\frac{D^{A}\rho_{(n)}}{\rho_{(n)}}  
- \frac{1}{2} 
\sqrt{\frac{r-2m}{r^3}} 
\frac{ \mathcal{S}^{A}}{\rho_{(n)}} - \frac{1}{2 \rho_{(n)}}\frac{m}{\sqrt{r^3(r-2m)}}\,  \mathcal{S}^{A}\, .
\end{align}
This is the EM memory vector for a Lagrangian observer in Schwarzschild space-time, regarding which we want to mention the following points: First, 
in the limit, $r \rightarrow \infty$, reduces to Minkowski space-time expression \eqref{eq:memory_minkowski}. Second, in the limit $r \rightarrow \infty$, the subleading term is proportional to $r^{-1}$. 
Third, to derive the memory vector $\Delta u^{A}$, we have to switch to the null time coordinate $u = t -r$ and integrate Eq.~\eqref{eq:memory_vec_kerr_2space_sim} with respect to $u$ at the asymptotic limit. 
Lastly, to evaluate the memory effect experienced by the static asymptotic (Lagrangian) observer, we need to do the 
transformation from $\left(\Vec{e}^{\, \theta}, \Vec{e}^{\, \phi}\right)$ to the orthogonal coordinate basis $\left(\hat{\theta}, \hat{\phi}\right)$ like in Sec.~\eqref{subsec:memory_Minkowski}.


\section{Conclusions} 
\label{sec:conclusion}

In this work, we have derived a master equation for electromagnetic memory in an arbitrary space-time. We used the covariant formalism to obtain the same. More specifically, we used the $1+1+2$ {covariant} formalism. The $1 + 1 + 2$ decomposition of space-time is a natural extension of the $1 + 3$ formalism in which the three-space is further decomposed using a given spatial direction. This choice of covariant formalism is because the net momentum (kick) vector lies on the 2-D surface for arbitrary space-time. Also, the electric and magnetic fields are transverse to the direction of propagation of the passing EM wave.

The EM memory \eqref{eq:memory_generic_w/o_orthogonality_final} has three distinct contributions: First contribution is due to the change in the EM field energy ($\rho_{(n)}$) along $n^{\mu}$ in the 2-space. This is non-zero for Minkowski space-time. 
The second contribution is proportional to the energy flux ($S^{\alpha}$) in the 2-space. This has kinematical information of the space-time and vanishes for the flat space-time. 
The third contribution is proportional to the acceleration ${\cal A}$ along the time-like vector $u^{\mu}$. To our understanding, the earlier approaches could not isolate the different contributions to the EM memory as done in this work.

Thus, the master equation \eqref{eq:memory_generic} corresponding to the acceleration of the comoving observer in the 2-D surface provides a physical understanding of the contribution to the memory. For instance, the leading order contribution only requires information about the total energy density of the EM field, while the subleading contributions contain information about the space-time geometry and the other components of the energy-momentum tensor of the EM field. This is the first time a transparent and easily applicable final expression for electromagnetic memory has been derived for a general curved space-time. Note that derivation of the master equation \eqref{eq:memory_generic} does not rely on the asymptotic properties. The analysis only requires the notion of comoving observers. This contrasts with the earlier works where one needs to assume a specific asymptotic nature of the fields and space-time.

We then obtained the EM memory for different space-times. In the case of FLRW space-time, we showed that the earlier analysis did not account for the geometric contribution to the memory effect~\cite{2022-EnriquezRojo-JHEP}. Specifically, their analysis did not account for the geometric contribution leading to a non-zero energy flux ($\mathcal{S}^{A}$) contribution. We have 
also obtained the EM memory for Kerr space-time. We also showed that the EM memory has a distinct signature for different pp wave space-times and can potentially be used as a probe.

It would be interesting to extend our analysis for black holes with multiple horizons and those that are not asymptotically flat. These may be particularly relevant for using EM memory as a probe to PBH. Finally, our analysis points to the possibility of using $1+1+2$ {covariant} formalism to understand gravitational memory in a unified manner~\cite{2022-Ferko.etal-PRD}. These are currently under investigation.

\begin{acknowledgements}
The authors thank A. Chowdhury, S. Mahesh Chandran and A. Kushwaha for comments on the earlier version of the manuscript and S. Maharaj regarding the $1 + 1 + 2$ formalism. The work is supported by the SERB-Core Research grant. 
\end{acknowledgements}


	

\providecommand{\href}[2]{#2}\begingroup\raggedright\endgroup

\end{document}